\pdfoutput=1
\documentclass[showpacs,prb]{revtex4}

\usepackage{amsmath}
\usepackage{amssymb}
\usepackage{color}
\usepackage{graphics}
\usepackage{graphicx}
\usepackage{dcolumn}

\renewcommand\Re{\operatorname{Re}}

\begin{document}

\title{Giant Faraday rotation due to excitation of magnetoplasmons in graphene microribbons}

\author{M. Tymchenko}
\affiliation{A. Ya. Usikov Institute for Radiophysics Electronics, NAS of Ukraine, 61085 Kharkiv, Ukraine}
\affiliation{Instituto de Ciencia de Materiales de Arag\'{o}n and Departamento de F\'{\i}sica de la Materia Condensada, CSIC-Universidad de Zaragoza, 50009 Zaragoza, Spain}
\author{A. Yu. Nikitin}
\affiliation{CIC nanoGUNE Consolider, 20018 Donostia-San Sebasti\'{a}n, Spain}
\affiliation{Ikerbasque, Basque Foundation for Science, 48011 Bilbao, Spain}
\author{L. Mart\'{\i}n-Moreno}
\email{lmm@unizar.es}
\affiliation{Instituto de Ciencia de Materiales de Arag\'{o}n and Departamento de F\'{\i}sica de la Materia Condensada, CSIC-Universidad de Zaragoza, 50009 Zaragoza, Spain}

\begin{abstract}
    A single graphene sheet, when subjected to a perpendicular static magnetic field provides
    Faraday rotation that, per atomic layer, greatly surpasses that of any other known material.
    This Giant Faraday rotation originates from the cyclotron resonance of massless electrons, which allows dynamical
    tuning through either external electrostatic or magnetostatic setting.
    Furthermore, the rotation direction can be controlled by changing the sign of the carriers in graphene,
   which can be done by means of an external electric field. However, despite these tuning possibilities, the
    requirement of large magnetic fields hinders application of the Faraday
    effect in real devices, especially for frequencies higher than few THz.
    In this work we demonstrate that, for a given value of the static external magnetic field,
    giant Faraday rotation can be achieved in arrays of graphene microribbons
    at frequencies much higher than the corresponding cyclotron frequency.
    The main feature in the magneto-optical response of graphene ribbons is not associated with the
    cyclotron resonance but rather with the fundamental magnetoplasmon excitation of a single ribbon.
    The magnetoplasmon nature of
    Faraday rotation in graphene ribbons opens great possibilities, as the resonance frequency
    can be locally selected by appropriately choosing the width of the ribbon while still preserving the
    tuning capability through a (smaller) external magnetic field.
\end{abstract}

\maketitle

The polarization of a plane wave is rotated when light passes through a transparent media in the presence of a
perpendicular static magnetic field $B$. This phenomenon, known as Faraday rotation, has important applications in optical diodes \cite{Shamir2006}, sensing and magnetic microscopy \cite{Hopster2005}, etc.
Recently, it has been found that a single graphene sheet provides a giant Faraday angle ($\sim 6^o$ at $B=7T$)\cite{Crassee2010}, which has given a strong impulse for
further investigations on graphene exposed to an external magnetic field. On top of its potential practical applications,
Faraday rotation is a powerful practical tool for studying the intrinsic properties of graphene samples.

Faraday rotation in a 2D graphene sheet arises from the excitation of the cyclotron resonance,
originated from the circular motion of conducting electrons. Even though the cyclotron frequency
is much larger for Dirac fermions than for massive ones, large Faraday rotation in graphene
even at THz frequencies still requires magnetic fields of the order of a few Teslas.
It would be desirable both to extend the spectral range to higher frequencies and to lower the
magnetic field at which substantial rotation occurs.
It has been demonstrated that opening periodically small gaps in a graphene sheet introduces capacitive effects,
which modify the effective impedance of graphene, providing an additional (geometrical) handle on the Faraday effect\cite{Fallahi2012}.

A related but different possibility, which will be explored in this paper, is to involve the resonant excitation of collective electron modes known as surface plasmons.
At zero magnetic field, this process strongly enhances the coupling between graphene
and incident radiation.\cite{Ju2011, Nikitin2011, Nikitin2012, Fang2012, Gao2012, Vasic2013, Fang2013}.
In the presence of a static magnetic field, applied perpendicularly to graphene,  plasmons and cyclotron excitations hybridize,
leading to the formation of graphene magnetoplasmons (GMP).\cite{Chiu1974, Kukushkin2006, Bychkov2008, Berman2008, Ferreira2012}
These GMP modes are known to significantly modify  the magneto-optical response of
graphene structures.\cite{Sounas2011, Fischer2011, Balev2011, GomezDiaz2012}.

Here we show that patterned graphene can be used to obtain plasmon-assisted giant Faraday rotation.
It must be noted that similar schemes have been studied in the optical regime by combination of a plasmonic grating and a magneto-optically active thin film (see e.g. \cite{Giessen2013}).
Graphene plasmonic gratings, which combine giant Faraday rotation in graphene with the strong coupling between radiation and graphene plasmons,
may provide an exciting prospect of dynamically tunable ultra-thin devices in both THz and infrared regimes, by employing solely the magneto-optical properties of graphene.

%

\begin{figure}
    \includegraphics{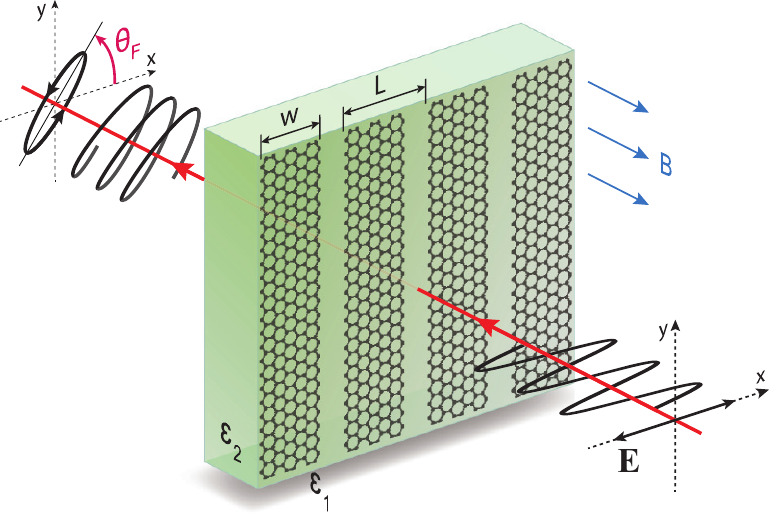}
    \caption{\label{Fig1} Schematic representation of the studied system:  a plane monochromatic wave is normally
     incident on the array of graphene ribbons, in the presence of a static perpendicular
     magnetic field $B$. $\theta_{\text{F}}$ is the Faraday rotation angle.}
\end{figure}

In this paper, for proof of principle, we consider a periodic array of graphene ribbons (with the width $w$ and period $L$),
illuminated at normal incidence by a monochromatic plane wave. A static magnetic field $B$ is
applied perpendicularly to the ribbons (see Fig.~\ref{Fig1} for a scheme of the considered geometry).
The incident electric field is chosen to lie perpendicularly to the axis of the ribbon (p-polarization). In this way, any deviation in the direction of the transmitted electric field is due to the effect in the magnetic field, and not to a "filtering" effect arising from the different transmittances for s- and p-polarized waves, which would occur even at zero field.~\cite{BludovJAP012}

Graphene is represented by its semi-classical conductivity tensor:\cite{Ferreira2011}
\begin{equation}\label{1}
    \begin{split}
        & \sigma_{xx} =  \sigma_{yy}  = \frac{e^2 |E_\text{F}|}{\hbar^2 \pi}\frac{i(\omega + i/\tau)}{(\omega + i/\tau)^2 - \omega_{\text{c}}^2}, \\
        & \sigma_{xy} =  - \sigma_{yx} = \frac{e^2 |E_\text{F}|}{\hbar^2 \pi}\frac{\omega_{\text{c}}}{(\omega + i/\tau)^2 - \omega_{\text{c}}^2},
    \end{split}
\end{equation}\label{2}
where $E_\text{F}$ is a Fermi energy, $\tau$ is a relaxation time,
and $\omega_{\text{c}}=eBv_\text{F}^2/|E_\text{F}|$ is the cyclotron frequency.
Here $v_\text{F}$ is the Fermi velocity of the Dirac fermions in graphene
($v_\text{F}=9.5\cdot10^5$~m/s).
Throughout this paper we consider that the system is at room temperature $T=300$~K,
and take a the representative value for the Fermi energy $E_\text{F}=0.2$~eV.
Our main conclusions are independent
on this choice, but it must be noticed that
the semiclassical expression for the conductivity incorporates only
intraband transitions\cite{Ferreira2011, Gusynin2009, Orlita2008}, {\it i.e.} is thus valid
for frequencies $\hbar\omega < 2 E_\text{F}$.
The spectral range we study in this paper is consistent with this restriction.
We will consider two values for the  relaxation time: $\tau=0.1$~ps  (corresponding to the mobility of about
$4500~\text{cm}^2 \text{V}^{-1} \text{s}^{-1}$) which is typical for
experimentally studied graphene ribbons, and $\tau\simeq4.4$~ps
(corresponding to the mobility of $200~000~\text{cm}^2 \text{V}^{-1} \text{s}^{-1}$) which is expected to be
achievable for suspended graphene at room temperatures if intrinsic
disorder is eliminated.\cite{Morozov2008, Hwang2008, Chen2008}

In order to find the scattering coefficients, we expand the total electromagnetic field in
the conventional form of the Fourier-Floquet plane waves expansion, in both upper and lower semi-spaces.
Matching appropriately the fields at the ribbon array yields an infinite set of equations
for the amplitudes of these waves. Details of the method can be found in the Supplementary Information.
The numerical calculation the amplitudes from the truncated system converges slowly
(and what is worse, non-uniformly)  with the number of diffraction orders considered.
We have validated that results obtained by this method have converged by comparing them with those obtained
by finite element calculations\cite{comsol} (which are more time consuming),
performed regularly for representative sets of parameters.
When only the zero-order diffraction mode is radiative (i.e, when $L<\lambda$,
which is the situation analyzed in this paper), the Faraday rotation angle can be
computed from the amplitudes on the $s$ and $p$ transmitted plane waves,  $t_{0xx}$ and $t_{0xy}$, respectively, as:\cite{Fialkovsky2009}

\begin{equation}\label{4}
   \theta_{\text{F}} =  \frac{1}{2} \arg \frac{t_{0xx} - i t_{0xy}}{t_{0xx} + i t_{0xy}}.
\end{equation}

For zero magnetic field, the dependence of transmission, reflection, and absorption spectra of
such structures on both period and width of the ribbons has been extensively studied theoretically
and experimentally.\cite{Ju2011,Nikitin2012, Alaee2012, Gao2012,IBM012}
In particular, it was found that the main resonance in the scattering coefficients is associated with the excitation of
a hybridized mode, which is a linear combination of the two edge modes of the ribbon.\cite{Nikitin2012, Eliasson1986, Mikhailov2005}
Additionally, there also exists an infinite set of weaker resonances which emerge due to coupling to waveguide-like graphene plasmons.
As we will discuss below, in a perpendicular magnetic field the ribbon plasmon modes transform into magnetoplasmon excitations but the number of modes and their field structure remains the same.

\begin{figure}[!t]
    \includegraphics{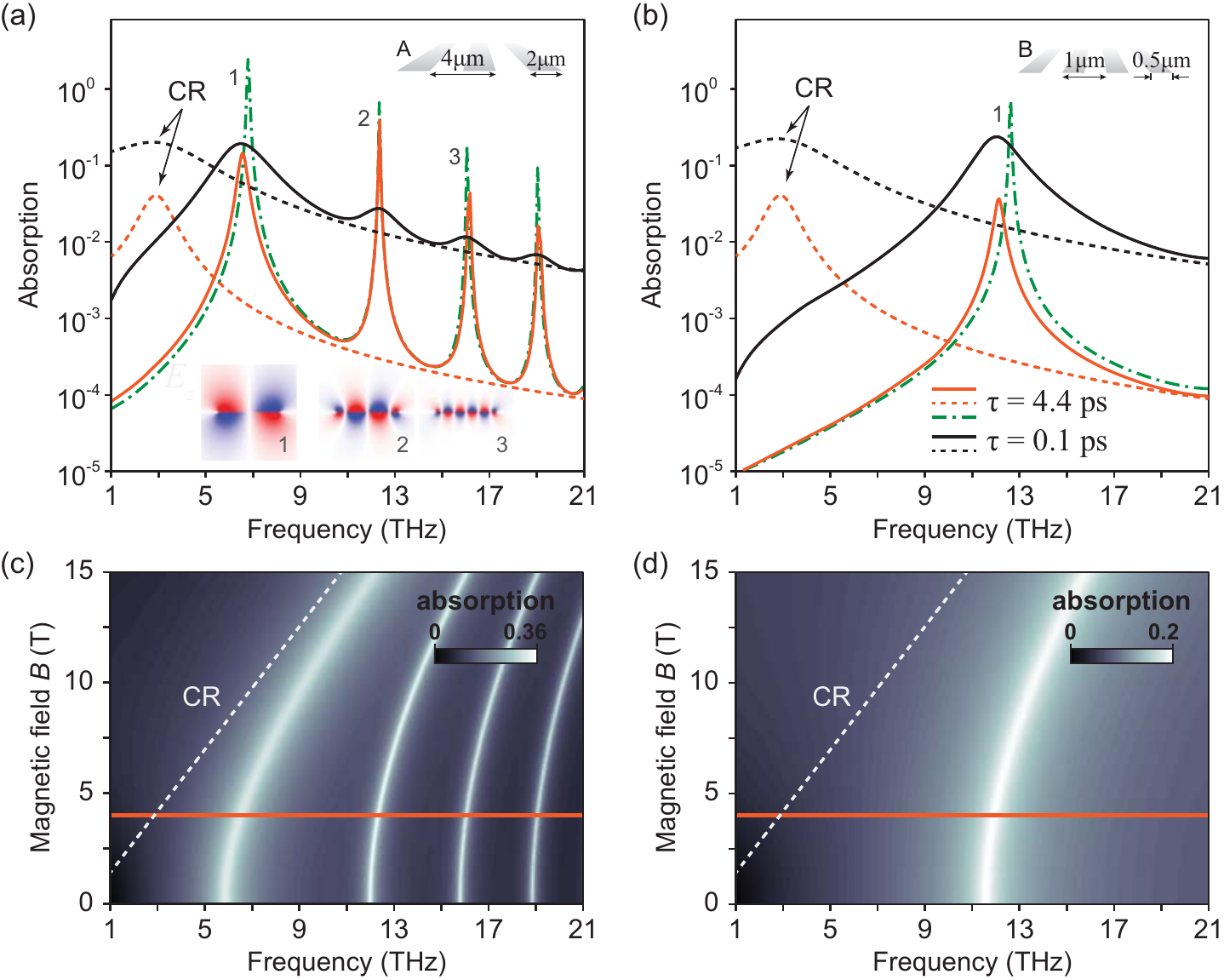}
    \caption{\label{Fig2} Panels (a) and (b): Absorption spectra of two different ribbon arrays (continuous lines),
    with geometrical parameters specified in each panel, and a uniform graphene sheet (dashed curves),
    in a magnetic field $B=4~\text{T}$. Two different relaxation times have been considered:
    $\tau = 4.4$~ps (orange curves) and  $\tau = 0.1$~ps (black curves).
    Green dot-dashed curves show the absorption cross-section for a single ribbon, for $\tau = 0.1$~ps.
    The structure of $E_z$ at the first three resonances indicated with ``1'', ``2'', ``3'' is shown in the inset of (a). Panels
    (c) and (d) show the absorption of the arrays A and B (geometrical parameters specified in panels (a) and (b), respectively) as a function of both frequency and magnetic field $B$.
    White dashed lines indicate the position of the cyclotron resonance (CR) in a uniform graphene sheet.
    Orange solid lines indicate the position of the ``cross-sections'' shown in panels (a) and (b).}
\end{figure}

In what follows we study the magnetoresponse of two structures, representative of those used in recent experiments (performed at $B=0$).\cite{Ju2011}: array ``A'', defined by ribbon width $w = 2~\mu\text{m}$ and period $L=4~\mu\text{m}$ and
array ``B'', with $w = 0.5~\mu\text{m}$ and $L=1~\mu\text{m}$.

Let us first analyze the absorption of radiation by the free-standing
ribbon arrays ($\varepsilon_1 = \varepsilon_2 = 1$).
In Fig.~\ref{Fig2} (a) and (b) and for the case $B=4~\text{T}$, we show the absorption for a continuous graphene sheet and the arrays considered. This figure also renders the absorption cross-section of the corresponding single free-standing graphene ribbons showing that, for the geometrical parameters considered, the absorption resonances are essentially due to the GMPs of  individual ribbons. In the array, the inter-ribbon coupling results only in a slight red-shifting of the main absorption line. In the considered frequency range, where the conductivity is given by a Drude term, the ribbon GMP modes are similar to those studied in stripes in two-dimensional electron gases arising in GaAs heterostructures.
\cite{Eliasson1986, Demel1988, Demel1991, Zhao1994, Mikhailov2005, Kukushkin2005, Kukushkin2006, Mikhailov2007, Fedorych2009}
Those studies showed that the frequencies of the magnetoplasmon modes $\omega_\text{n}(B)$ are given by a simple expression
\begin{equation}\label{3}
    \omega_\text{n}(B) = \sqrt{\omega_{\text{n}}^2 + \omega_{\text{c}}^2}.
\end{equation}
where $\left\{\omega_{\text{n}}\right\}$ is the set of plasmon frequencies in the ribbon at zero magnetic field.
Each eigenfrequency $\omega_{\text{n}}$ scales with the ribbon width as
$\sqrt{n/w}$, where $n$ is a number of the mode, $n= 1, 2, 3,...$.\cite{Ju2011, Mikhailov2005}
(in the considered geometry, only the odd orders are excited).
The dependence of the absorption spectra with DC magnetic field for both arrays is rendered in Fig.~\ref{Fig2} (c) and (d),
reflecting clearly the evolution of the GMP frequencies.

\begin{figure*}[!t]
    \includegraphics{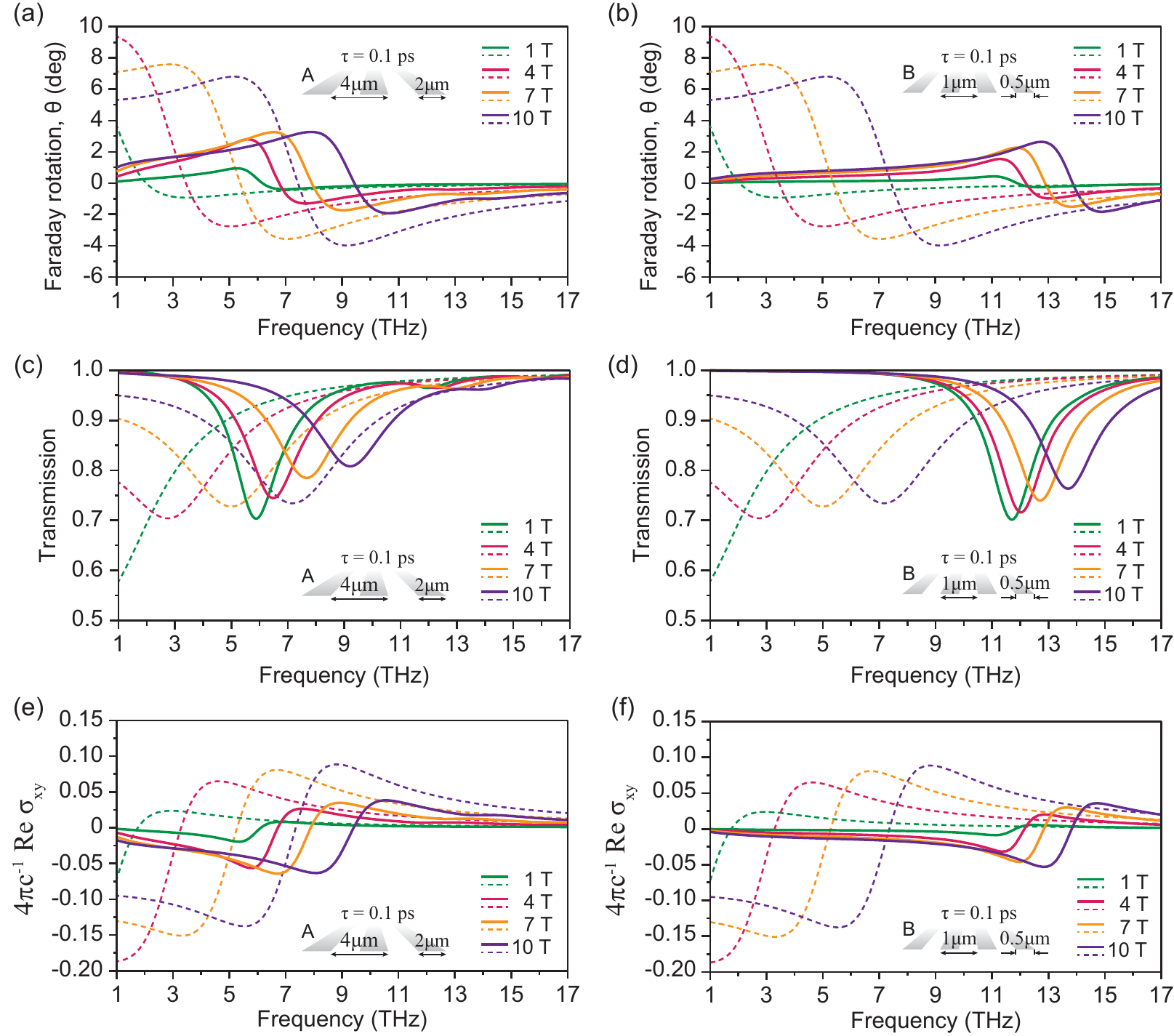}
    \caption{\label{Fig3} The results for ribbon arrays (solid curves) and continuous graphene (dashed curves) for
    relaxation time $\tau = 0.1$~ps. Panels to the left correspond to array "A" ($w=2~\mu\text{m}$, $L=4~\mu\text{m}$) while right panels
    are for array "B" ($w=0.5~\mu\text{m}$, $L=1~\mu\text{m}$).
   Upper panels: Faraday rotation angle at various magnetic fields.
   Middle panels: Zeroth-order transmission $|t_0|^2 = |t_{0xx}|^2+|t_{0xy}|^2$.
   Lower panels: The real part of the non-diagonal component of the effective conductivity, $\Re\widetilde\sigma_{xy}$ (solid curves),
    and   $\Re\sigma_{xy}$ (dashed curves).}
\end{figure*}

Panels (a) and (b) in Fig.~\ref{Fig3} show the spectra for the Faraday rotation angle for the
two ribbon arrays considered, and for different DC magnetic fields (solid lines). For comparison,
in the same panels we present the Faraday rotation angle
for a continuous graphene sheet (dashed lines).
As mentioned above, in continuous graphene the main resonance feature is associated
to the cyclotron resonance. Therefore,
giant Faraday rotation is restricted to rather low frequencies, less than $10$~THz,
even at high magnetic fields (see white dashed lines in lower panels in Fig.~\ref{Fig2}).
As seen from Fig.~\ref{Fig3}, graphene ribbon arrays,
despite being a diluted one-atom thick material, still present values of $\theta_{\text{F}}$ of the order of a few degrees.
Moreover, the frequency at which maximum $\theta_{\text{F}}$ occurs is, in ribbon arrays,
are strongly blue shifted with respect to the one in a continuous graphene sheet.
This difference is especially significant for low magnetic fields (less than a few Tesla). In the
ribbon array, the maximum of Faraday rotation occurs at the
the resonance excitation of GMPs, which
is determined by both ribbon width and magnetic field.
At high magnetic fields, for which the fundamental GMP mode approaches the cyclotron resonance,
the ribbon arrays have no obvious advantage over continuous graphene, producing even a slightly weaker
Faraday effect.

Fig.~\ref{Fig3}(c) and (d) show the corresponding transmittance spectra.
Notice that minimum transmittance occurs approximately at the spectral position when $\theta_{\text{F}}$ changes the sign.
Importantly,  the transmittance is at maximum $\theta_{\text{F}}$ is of the order of $\sim 85-90\%$,
which allows increasing of the Faraday rotation by stacking vertically
several layers of ribbon arrays, yet maintaining an appreciable transmitted field.

When the lattice parameter is much smaller than the free-space wavelength we can use a metamaterial approach and, for the computation of the scattering coefficients, represent the graphene ribbon array as a continuous monolayer
with an effective conductivity tensor $\widetilde\sigma(B)$.
The two independent components of the effective conductivity tensor,
namely $\widetilde\sigma_{xx}$ and $\widetilde\sigma_{xy} $, can be uniquely derived from the zero-order
transmission coefficients $t_{0xx}$ and $t_{0xy}$.
In particular, for the free-standing ribbons array we obtain (see Supplementary Information for the derivation):
\begin{equation}\label{5}
    \begin{split}
        &\widetilde\sigma_{xx} = \frac{c}{4\pi}\left[\frac{t_{0xx}}{t_{0xx}^2+t_{0xy}^2} - 1\right],\\
        &\widetilde\sigma_{xy}=  \frac{c}{4\pi} \, \frac{t_{0xy}}{t_{0xx}^2+t_{0xy}^2}.
    \end{split}
\end{equation}

The lower panels of Fig.~\ref{Fig3} show the frequency dependence of both $\Re\widetilde\sigma_{xy}$ (for the two
considered arrays) and $\Re\sigma_{xy}$ (continuous graphene). Notice that the relation
between the Faraday angle and non-diagonal component of the conductivity tensor, valid for low relaxation times,
$\theta_{\text{F}} \simeq 4\pi c^{-1} \Re \sigma_{xy}$ can be still applied in the case of ribbon arrays (in which case $\theta_{\text{F}} \simeq 4\pi c^{-1} \Re \widetilde\sigma_{xy}$), as can be seen by comparing Fig.~\ref{Fig3}(e) and (f) with  Fig.~\ref{Fig3}(a) and (b).

\begin{figure}[!t]
    \includegraphics{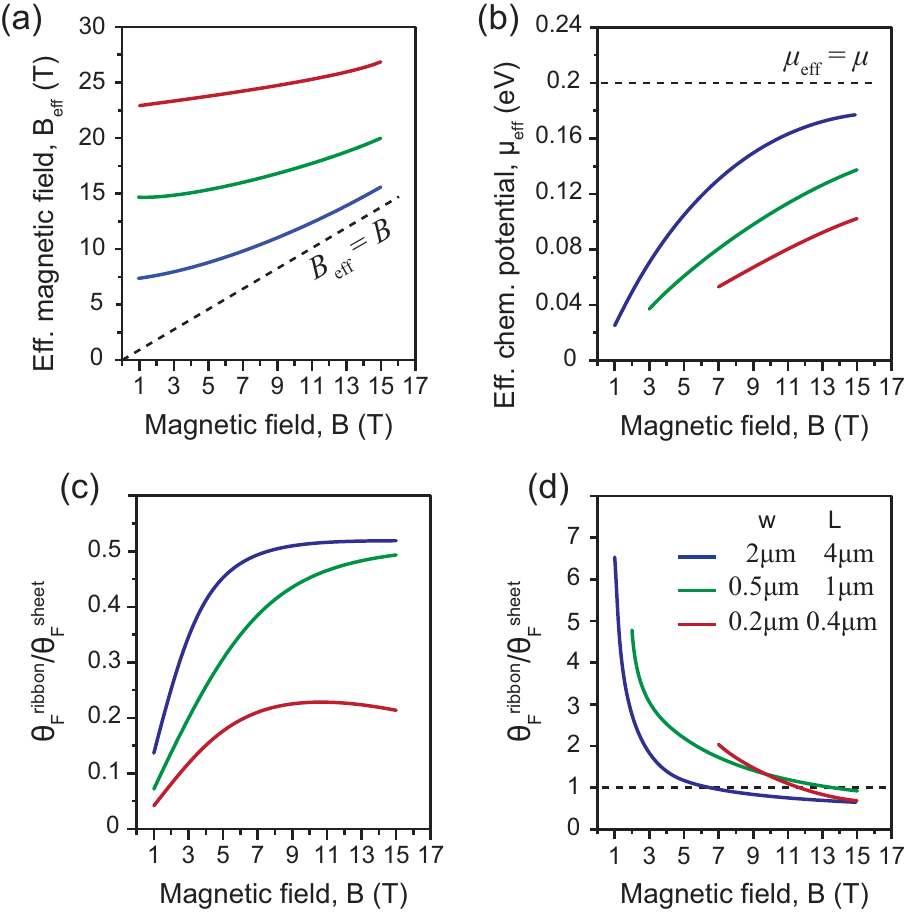}
    \caption{\label{Fig4} Effective magnetic field (a) and effective chemical
    potential (b), defined as those at which a continuous graphene sheet
  would produce maximum giant Faraday rotation at the same frequency as
  the ribbon array at a given magnetic field $B$ (with other parameters unchanged).
  The bottom panels render the ratio between the maximal Faraday angle of
  graphene ribbon arrays and that of the
  continuous sheet at the corresponding effective parameters (panel (c) for
$B_{\text{eff}}$ and panel (d) for $\mu_{\text{eff}}$).}
\end{figure}

In order to better illustrate the magneto-optical response of the ribbon arrays,
we define the effective magnetic field, $B_{\text{eff}}$, such that the spectral
position of the maximum Faraday angle
is the same for a {\it continuous}
graphene sheet under the presence of $B_{\text{eff}}$ and for the ribbon array
under the actual $B$ (with all other parameters, like $T$, $\mu$, etc., remaining the same).
Fig.~\ref{Fig4} (a) renders the dependence of the computed $B_{\text{eff}}$ with $B$,
for different arrays of ribbons, showing clearly that sub-micrometer graphene ribbons
produce maxima in the Faraday rotation at frequencies that would only be achievable
in a continuous graphene sheet at much larger magnetic fields. If these large magnetic field
are achievable, a continuous sheet would produce a larger Faraday angle.
This is illustrated in Fig.~\ref{Fig4} (c),
which shows the ratio between $\theta_{\text{F}}(B_{\text{eff}})$ in a graphene sheet
and $\theta_{\text{F}}(B)$ in the ribbon, $R(B)=\theta_{\text{F}}^{\text{ribbon}}(B)/ \theta_{\text{F}}^{\text{sheet}}(B_{\text{eff}})$, evaluated at the resonant frequency.
This figure shows that the decrease in magnetic field needed to obtain
high Faraday angles in ribbons, with respect to the situation in a continuous sheet, comes at a decrease in the maximum Faraday angle ($R<1$). As the magnetic field increases, so does $R$ and,
when the magnetic field is high enough (so that the magnetoplasmons frequency
in ribbons tend to the cyclotron frequency), $R(B)$ is limited by the filling fraction $w/L$.

Another way to obtain Faraday rotation at higher frequencies in a continuous graphene sheet
is by reducing the chemical potential. In order to compare the response of 2D graphene and
the ribbon array, we define an effective chemical potential $\mu_{\text{eff}}(B)$ such
that the spectral position of maximum $\theta_{\text{F}}$ is the same for a {\it continuous}
graphene sheet at $\mu_{\text{eff}}(B)$ and for the ribbon array
under the actual $\mu$ (with all other parameters, like $T$, $B$, etc., remaining the same).
Figure~\ref{Fig4} (b) renders the  $\mu_{\text{eff}}(B)$ for several ribbons arrays, for $\mu=0.2~\text{eV}$. Notice
however, that the strategy of increasing the frequency of resonant Faraday rotation in a continuous sheet by decreasing the chemical potential is limited by the condition $\hbar \omega_{c}<2 \mu_{\text{eff}}$. In Fig.~\ref{Fig4} (c) this is reflected in the fact that the different curves have end-points at the lower chemical potentials that fulfill the previous condition. Notice also that, as shown in Fig.~\ref{Fig4} (d), ribbon
arrays can provide, at a given magnetic field and a given frequency, a substantially larger Faraday rotation than a continuous graphene sheet. This occurs at magnetic fields such that $\mu_{\text{eff}}<<\mu$, i.e., when the resonance is magnetoplasmonic, rather than cyclotronic, in character.

\begin{figure}[!t]
    \includegraphics{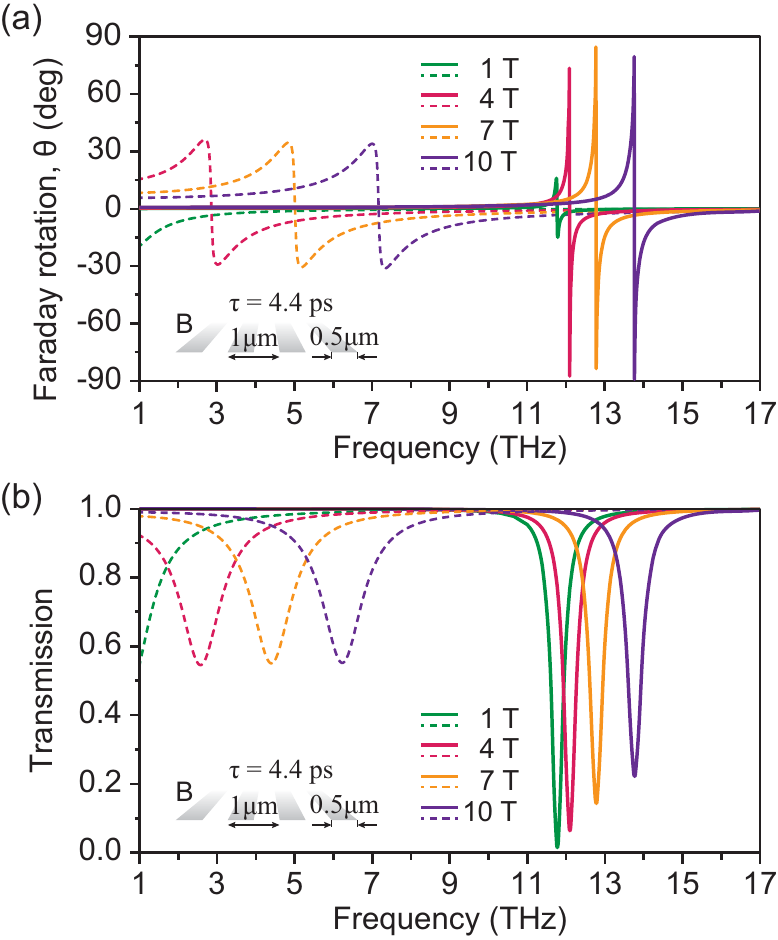}
    \caption{\label{Fig5} The results for ribbon arrays (solid curves) and continuous graphene (dashed curves) for the ultrahigh relaxation time $\tau = 4.4$~ps.
    (a) Faraday rotation at various magnetic fields. (b) Zeroth-order transmission.}
\end{figure}

Up to now we have considered a value for the carrier mobility in graphene that is
routinely produced nowadays. However, in view of the recent advances in
producing high-quality graphene samples, it is interesting to consider which Faraday rotation
would be achievable at ultrahigh mobilities.
For that, it is valuable to know whether there are physical limits to what can be achievable. A well-known example of these limits is that the maximum possible absorption by a
free standing infinitely thin sheet is $50\%$\cite{Suko2012}.
Concerning the change in polarization, we find that in a free-standing infinitely thin sheet,
characterized by a local conductivity, the modulus each of the zero-order cross-polarization amplitudes
have a maximum value of 1/2 (see Supplementary Information). This is,
\begin{equation}\label{ni18}
 |t^{sp}|_{\text{max}}^2 = |t^{ps}|_{\text{max}}^2 =  |r^{sp}|_{\text{max}}^2 = |r^{ps}|_{\text{max}}^2 =  \frac{1}{4}.
\end{equation}
By structuring graphene, we cannot increase this limiting value. However,
we can largely improve the resonance q-factor (the GMP resonance
is less absorptive, see Fig.~\ref{Fig2}(a),(b)). This has implications for the Faraday rotation.
In a continuous translational-invariant sheet it can
be shown ((see Supplementary Information) that the maximal Faraday angle is
$|\theta_{\text{F}}|_{\text{max}} \simeq \pi/4$ which, of course, is
already an impressive result, even more so for a one-atom-thick layer. Remarkably, Faraday rotation on graphene ribbon arrays can exceed this value.
As was previously shown,\cite{Nikitin2012} the transmission coefficient $t_{0xx}$ be very low.
If we set $t_{0xx}\rightarrow 0$ in Eq.~\eqref{4}, we immediately obtain $|\theta_{\text{F}}| \rightarrow \pi/2$.
These properties are illustrated in Fig.~\ref{Fig5}, which shows the computed transmittance
and Faraday angle, for both arrays A and B, for the case $\tau = 4.4$~ps.
This figure shows that the Faraday angle strongly depends on scattering time
and that in ribbons it may even exceed that of a continuous graphene sheet (and,
as stated above, at larger frequencies).

\begin{figure}[!t]
    \includegraphics{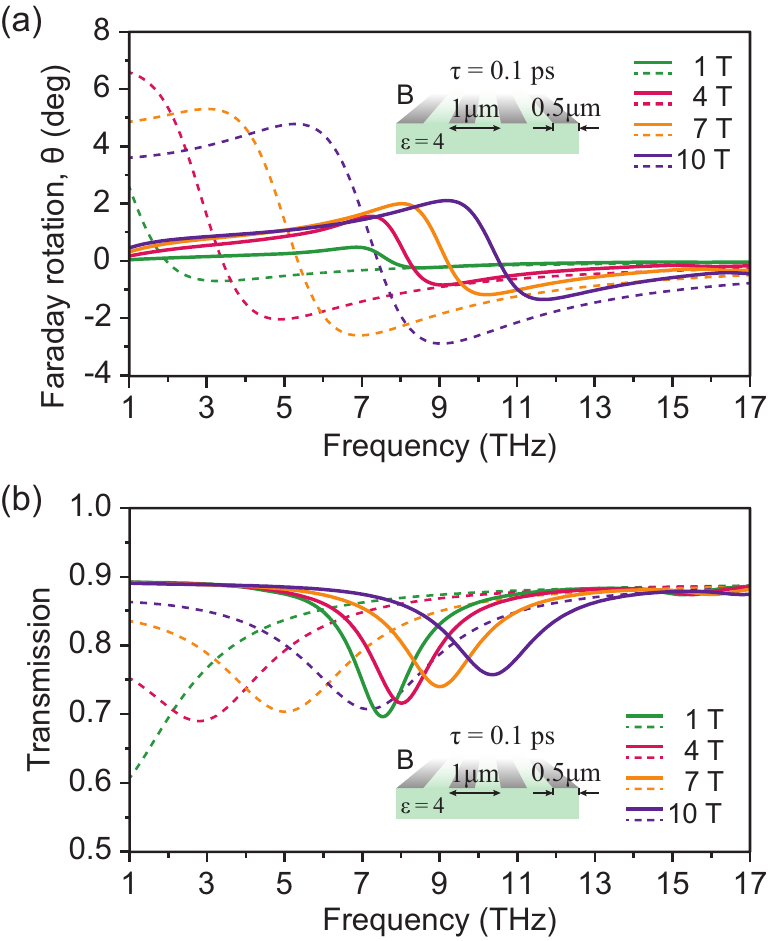}
    \caption{\label{Fig6} The results for ribbon arrays (solid curves) and continuous graphene (dashed curves) placed on top of the substrate with $\varepsilon = 4$.
     (a) Faraday rotation at various magnetic fields. (b) Zeroth-order transmission. The relaxation time is $\tau = 0.1$~ps.}
\end{figure}

In all the previous calculations, we have considered free-standing
graphene ($\varepsilon_1 = \varepsilon_2 = 1$) because the fundamental physics we were
discussing (the influence of excitation of GSP on the Faraday rotation) was
already present in this simple configuration.
Nevertheless, in practice, graphene is usually produced on a substrate. One
consequence of this is that carrier mobility is reduced, due to scattering with
phonons and charged impurities. Additionally, although
the presence of a substrate does not significantly affect the polarization rotation in a continuous graphene
sheet, this may not be the case for graphene ribbons, as plasmon excitations are very
sensitive to the surrounding medium. The influence of a substrate on the GSP-enhanced
Faraday rotation of graphene ribbons is exemplary illustrated in Fig.~\ref{Fig6}, for one
of the ribbon arrays considered in this work placed on a semi infinite substrate characterized
by a non-dispersive dielectric permittivity $\varepsilon_2  = 4$.
As the figure shows, the blue shifts of the maximum in the Faraday angle occurring when
graphene is patterned are still present, although they are reduced by the presence of the
dielectric substrate (as can be seen by the comparison with Fig~\ref{Fig3}(b)).

In conclusion, we have shown that, for a given magnetic field and chemical potential, structuring graphene
periodically can produce strong Faraday rotation at larger frequencies than
what would occur in a continuous graphene sheet. Alternatively, at a given frequency, graphene ribbons produce giant Faraday rotation at much smaller magnetic fields than in continuous graphene. We have demonstrated that this enhancement of Faraday rotation in arrays of graphene ribbons is produced by the resonant coupling of radiation to graphene  magnetoplasmons.
The possibility to control the graphene magnetoplasmon
frequency through both geometry and magnetic field, combined with the
possibility to modify the carrier density by an external gate voltage, holds an exciting promise for designing dynamically tunable devices employing solely the
magneto-optical properties of graphene.

\newpage

Supplementary information

\section{Conductivity model}

In an external magnetic field, due to the cyclotron motion of charge carriers, the conductivity tensor acquires non-diagonal components
\begin{equation}\label{1}
    \hat{\sigma} = \left( \begin{matrix} \sigma_{xx} & \sigma_{xy} \\ \sigma_{yx} & \sigma_{yy} \end{matrix}\right)
    =  \left( \begin{matrix} \sigma_{1} & \sigma_{2} \\ -\sigma_{2} & \sigma_{1} \end{matrix}\right),
\end{equation}
In case of rather low energies, $\hbar \omega < 2 \mu$, the main contribution to the graphene conductivity arises for intraband transitions.
For such energies the Boltzmann's transport theory is valid, so that neglecting temperature effects in the Fermi distribution, the two independent components ($\sigma_1$ and $\sigma_2$)
of the conductivity tensor can be expressed as:
\begin{equation}\label{1a}
   \sigma_{1} = \frac{e^2}{\hbar \pi}\frac{i (\Omega+i\gamma)}{ (\Omega+i\gamma)^2 - \beta^2}, \quad
   \sigma_{2} = \frac{e^2}{\hbar \pi}\frac{ \beta}{ (\Omega+i\gamma)^2 - \beta^2}.
\end{equation}
Here $\Omega = \hbar\omega/\mu$, $\gamma = \mathcal{E}_{\text{s}}/\mu$, where $\mathcal{E}_{\text{s}} = \hbar/\tau$ is a scattering energy, and $\tau$ is a relaxation time.
The quantity $\beta$ is the cyclotron resonance energy normalized to the chemical potential
\begin{equation}\label{1b}
   \beta = \frac{\hbar\omega_c}{\mu}, \quad
   \omega_c = \frac{eBv_{\text{F}}^2}{|\mathcal{E}_{\text{F}}|},
\end{equation}
where $B$ is the magnetic field, $v_{\text{F}}\simeq9.5\cdot10^5~\text{m/s}$ is the Fermi velocity, and $\mathcal{E}_{\text{F}}$ is the Fermi energy that we set equal to the chemical potential, $\mathcal{E}_{\text{F}} = \mu$.

\section{Modal expansion}
Let as assume an arbitrary polarized monochromatic plane wave impinging onto a periodic array of graphene ribbons, in the presence of a static perpendicular magnetic field $B$.
We chose the coordinate system so that the ribbon array is located at $z=0$, between two semi-spaces dielectrics characterized by electrical permittivities $\varepsilon_1$ and $\varepsilon_2$, respectively.
 We chose the $x$ axis so that it coincides with the  conductivity modulation direction.
 Then, for the conductivity tensor we have $\hat{\sigma}(x) = \hat{\sigma}(x+L)$, where $L$ is
the array period. We present the conductivity tensor of the ribbons array as the Fourier expansion,
\begin{equation}\label{2}
    \hat{\sigma}(x) = \sum_n \hat{\sigma}_n e^{inGx}, \quad G = 2\pi/L,
\end{equation}
with the Fourier coefficients
\begin{equation}\label{2a}
    \hat{\sigma}_n = \frac{1}{L} \int_{-L/2}^{L/2} dx \hat{\sigma}(x) e^{-inGx}.
\end{equation}

At the patterned graphene layer, $z=0$, it is convenient to introduce the polarization unit-vectors and the corresponding plane wave eigenfunctions of the incident, reflected and transmitted waves.
For the incident wave the polarization unit vectors $\mathbf{s}^{i}$ and $\mathbf{p}^{i}$ can be presented as
\begin{equation}\label{2b}
    \mathbf{s}^{i} =  \frac{1}{k_t}\left( \begin{matrix} -k_y \\ k_x \\ 0 \end{matrix}\right), \quad
    \mathbf{p}^{i} = \frac{q_1}{k\sqrt{\varepsilon_1}k_t} \left( \begin{matrix} k_x \\ k_y \\ k_t^2/q_1 \end{matrix}\right),
\end{equation}
and the corresponding eigenfunctions
\begin{equation}\label{3}
    |s^i\rangle =  \mathbf{s}^{i} \cdot e^{ik_xx+ik_yy}, \quad
    |p^i\rangle = \mathbf{p}^{i} \cdot e^{ik_xx+ik_yy}.
\end{equation}
Here $k=\omega/c$, $k_t^2 = k_x^2+k_y^2$, and $q_{1} = \sqrt{\varepsilon_1 k^2 - k_t^2}$  so that $q_{1} > 0$.
To write the eigenfunctions of reflected and transmitted waves let us introduce tangential and normal to graphene components of their wavevectors
\begin{eqnarray}\label{4}
    \mathbf{k}_{nt} = \mathbf{k}_t + n\mathbf{G}, \quad  q_{1,2n} = \sqrt{\varepsilon_{1,2} k^2 - \mathbf{k}_{nt}^2}, \quad \text{Re,~Im}~q_{1,2n} \geq0,
\end{eqnarray}
with $\mathbf{G} = (G, 0, 0)$.
Then, the polarization unit vectors can be written as
\begin{equation}\label{5}
    \mathbf{s}^{a}_n  = \frac{1}{k_{nt}} \left( \begin{matrix} -k_y \\ k_{nx} \\ 0 \end{matrix}\right), \quad
    \mathbf{p}^{a}_n = \frac{q_{an}}{k\sqrt{\varepsilon_a}k_{nt}} \left( \begin{matrix} k_{nx} \\ k_y \\ -k_{nt}^2/k_{nz}^a \end{matrix}\right),
\end{equation}
so the corresponding plane wave eigenfunctions are
\begin{equation}\label{6}
    |s^a_n\rangle  = \mathbf{s}^{a}_n \cdot e^{ik_{nx}x+ik_yy}, \quad
    |p^a_n\rangle = \mathbf{p}^{a}_n \cdot e^{ik_{nx}x+ik_yy}.
\end{equation}
Here $a$ stands for ``$r$'' (reflected) and ``$t$'' (transmitted), and $k_{nz}^r \equiv q_{1n}$, $k_{nz}^t \equiv -q_{2n}$.
The electric fields in the upper and lower semi-spaces can be written as
\begin{equation}\label{7}
    \begin{split}
    \mathbf{E}_1 &= \sum_{\alpha = s,p} \left\{ E^{i,\alpha} |\alpha^i\rangle e^{-iq_{1}z} + \sum_n E_{n}^{r, \alpha} |\alpha_n^r\rangle e^{iq_{1n}z}\right\}, \\
    \mathbf{E}_2 &= \sum_{\alpha = s,p} \sum_n E_{n}^{t,\alpha} |\alpha_n^t \rangle e^{-iq_{2n}z},
    \end{split}
\end{equation}
where $\alpha = s,p$.
Let us also introduce the normalized in-plane polarization unit-vectors $\mathbf{e}^s_n$, $\mathbf{e}^p_n$:
\begin{equation}\label{7a}
    \mathbf{e}^s_n =  \frac{1}{k_{nt}} \left( \begin{matrix} -k_y \\ k_{nx} \\ 0 \end{matrix}\right), \quad
    \mathbf{e}^p_n =  \frac{1}{k_{nt}} \left( \begin{matrix} k_{nx} \\ k_y \\ 0 \end{matrix}\right),
\end{equation}
and in-plane modes $|s_n\rangle$, $|p_n\rangle$ which contain only tangential components of the fields
\begin{equation}\label{8}
    |s_n\rangle  =  \mathbf{e}^s_n \cdot  e^{ik_{nx}x+ik_yy}, \quad
    |p_n\rangle =  \mathbf{e}^p_n \cdot  e^{ik_{nx}x+ik_yy},
\end{equation}
and which do not depend on surrounding media since they do not contain $k_{nz}^a$. The relations between $|\alpha^i\rangle$, $|\alpha_n^a\rangle$ and $|\alpha_n\rangle$ are following:
\begin{equation}\label{9}
    |\alpha^i\rangle_t \equiv c_{1}^\alpha |\alpha_0\rangle, \quad
    |\alpha^r_n\rangle_t \equiv c_{1n}^\alpha |\alpha_n\rangle, \quad
    |\alpha^t_n\rangle_t \equiv c_{2n}^\alpha |\alpha_n\rangle,
\end{equation}
where $c_{1}^\alpha \equiv  c_{10}^\alpha$, and
\begin{equation}\label{10}
    c_{an}^s = 1, \quad c_{an}^p = \frac{q_{an}}{k\sqrt{\varepsilon_a}}.
\end{equation}

The boundary conditions at the patterned graphene layer $z=0$ read
\begin{equation}\label{11}
\left\{
    \begin{split}
    &\mathbf{E}_{1t} - \mathbf{E}_{2t} = 0, \\
    & \left[\mathbf{e}_z \times (\mathbf{H}_{1} - \mathbf{H}_{2})\right] = \frac{4\pi}{c}\hat{\sigma} \mathbf{E}_{2t}.
    \end{split}
\right.
\end{equation}
Substituting here the expressions for the fields (\ref7)  and taking into account that,
from Maxwell equations, it follows
\begin{equation}\label{12}
    \mathbf{H} = \frac{1}{k} \left[ \mathbf{k} \times \mathbf{E} \right],
\end{equation}
then, we obtain that the boundary conditions can be written as
\begin{equation}\label{13}
    \begin{split}
   & \sum_{\alpha = s,p} \left\{c_{1}^\alpha E^{i,\alpha} |\alpha_0\rangle + \sum_n \left(c_{1n}^\alpha E_{n}^{r,\alpha} - c_{2\alpha} E_{n}^{t,\alpha}\right) |\alpha_n\rangle \right\} = 0, \\
   & \frac{1}{k} \mathbf{e}_z \times \sum_{\alpha = s,p}
   \left\{ E^{i,\alpha} \mathbf{k}^i \times |\alpha^i\rangle + \sum_n \left(E_{n}^{r, \alpha} \mathbf{k}_n^r \times |\alpha_n^r\rangle  - E_{n}^{t,\alpha}  \mathbf{k}_n^t \times |\alpha_n^t\rangle \right) \right\} = \frac{4\pi}{c} \sum_m \hat{\sigma}_m e^{iGmx}\sum_{n, \alpha} E_{n}^{t,\alpha} |\alpha_n^t\rangle.
   \end{split}
\end{equation}
Introducing the surface impedances,
\begin{equation}\label{14}
    \begin{split}
    \frac{1}{k} \left[ \mathbf{e}_z \times \left[  \mathbf{k}^i \times |\alpha^i\rangle \right] \right] &= c_{1}^\alpha~Z_{1}^\alpha |\alpha_0\rangle,     \\
    \frac{1}{k} \left[ \mathbf{e}_z \times \left[  \mathbf{k}_n^r \times |\alpha_n^r\rangle \right] \right]  &= - c_{1n}^\alpha~Z_{1n}^\alpha  |\alpha_n\rangle,  \\
    \frac{1}{k} \left[ \mathbf{e}_z \times \left[  \mathbf{k}_n^t \times |\alpha_n^t\rangle \right] \right]  &= c_{2n}^\alpha~Z_{2n}^\alpha  |\alpha_n\rangle,
    \end{split}
\end{equation}
with
\begin{equation}\label{16}
     Z_{an}^s = \frac{q_{an}}{k}, \quad Z_{an}^p = \frac{k\varepsilon_{a}}{q_{an}},   \quad   Z_1^\alpha \equiv Z_{10}^\alpha,
\end{equation}
the boundary conditions become
\begin{equation}\label{17}
    \begin{split}
   & \sum_{\alpha = s,p} \left\{c_{1}^\alpha E^{i,\alpha} |\alpha_0\rangle + \sum_n \left(c_{1n}^\alpha E_{n}^{r,\alpha} - c_{2n}^\alpha E_{n}^{t,\alpha}\right) |\alpha_n\rangle \right\} = 0, \\
   &  \sum_{\alpha = s,p} \left\{ - c_{1}^\alpha E^{i,\alpha} Z_{1}^\alpha |\alpha_0\rangle + \sum_n \left(c_{1n}^\alpha E_{n}^{r,\alpha} Z_{1n}^\alpha + c_{2n}^\alpha E_{n}^{t,\alpha} Z_{2n}^\alpha \right) |\alpha_n\rangle \right\} =
   -\frac{4\pi}{c} \sum_m \hat{\sigma}_m e^{iGmx} \sum_{n,\alpha} E_{n}^{t, \alpha} c_{2n}^\alpha |\alpha_n\rangle.
   \end{split}
\end{equation}
Projecting these two equations on $\langle\alpha_n|$ and integrating the expression over the structure period we obtain:
\begin{equation}\label{18}
\left\{
    \begin{split}
   & c_{1n}^\alpha E^{i,\alpha} \delta_{n,0} +c_{1n}^\alpha E_{n}^{r,\alpha} - c_{2n}^\alpha E_{n}^{t,\alpha}  = 0, \\
   &  - c_{1n}^\alpha E^{i,\alpha} Z_{1}^\alpha \delta_{n,0} + c_{1n}^\alpha E_{n}^{r, \alpha} Z_{1n}^\alpha  + c_{2n}^\alpha E_{n}^{t, \alpha}  Z_{2n}^\alpha =
   -\frac{4\pi}{c} \sum_{m, \alpha'}  (\mathbf{e}^\alpha_n)^\mathrm{T} \hat{\sigma}_{n-m} \mathbf{e}^{\alpha'}_m  E_{m}^{t, \alpha'} c_{2m}^{\alpha'}.
    \end{split}
\right.
\end{equation}
Let us now define the transformation coefficients of the diffracted waves
\begin{equation}\label{19}
   E_{n}^{r, \alpha} = \sum_{\alpha'} r_{n}^{\alpha\alpha'} E^{i, \alpha'},   \quad
   E_{n}^{t,\alpha} = \sum_{\alpha'} t_{n}^{\alpha\alpha'} E^{i,\alpha'}.
\end{equation}
Substituting (\ref{19}) into (\ref{18}) we obtain
\begin{equation}\label{20}
\left\{
    \begin{split}
   & c_{1n}^\alpha \delta_{n,0}\delta_{\alpha,\alpha'} + c_{1n}^\alpha r_{n}^{\alpha\alpha'} - c_{2n}^\alpha t_{n}^{\alpha\alpha'} = 0, \\
   & - c_{1n}^\alpha Z_{1}^\alpha \delta_{n,0}\delta_{\alpha,\alpha'} + c_{1n}^\alpha r_{n}^{\alpha\alpha'} Z_{1n}^\alpha  + c_{2n}^\alpha t_{n}^{\alpha\alpha'}  Z_{2n}^\alpha  =
   - \sum_{m, \alpha''} f_{nm}^{\alpha\alpha''}  c_{2m}^{\alpha''} T_{m}^{\alpha''\alpha'},
    \end{split}
\right.
\end{equation}
where
\begin{equation}\label{21}
     f_{nm}^{\alpha\alpha'} = (\mathbf{e}^\alpha_n)^\mathrm{T} \hat{\alpha}_{n-m} \mathbf{e}^{\alpha'}_m, \quad
     \hat{\alpha}_{n-m} = \frac{4\pi}{c}\hat{\sigma}_{n-m} \equiv \left( \begin{matrix} \alpha_{1,n-m} & \alpha_{2,n-m} \\ -\alpha_{2,n-m} & \alpha_{1,n-m} \end{matrix}\right).
\end{equation}
Here, one should not confuse the components of the tensor $\hat\alpha_{n-m}$ with the polarization index $\alpha$.
From the first equation of \eqref{20} we have
\begin{equation}\label{23}
    t_{n}^{\alpha\alpha'} = \frac{c_{1n}^\alpha}{c_{2n}^\alpha}\left(\delta_{n,0}\delta_{\alpha,\alpha'} + r_{n}^{\alpha\alpha'}\right).
\end{equation}
Substituting  \eqref{23} into the second expression of \eqref{20} we obtain
\begin{equation}\label{24}
    - c_{1n}^\alpha Z_{1n}^\alpha \delta_{n,0}\delta_{\alpha,\alpha'} + c_{1n}^\alpha r_{n}^{\alpha\alpha'} Z_{1n}^\alpha  + c_{1n}^\alpha\left(\delta_{n,0}\delta_{\alpha,\alpha'} + r_{n}^{\alpha\alpha'}  \right) Z_{2n}^\alpha =
   - \sum_{m, \alpha''} f_{nm}^{\alpha\alpha''}  c_{1m}^{\alpha''} \left(\delta_{m,0}\delta_{\alpha''\alpha'} + r_{m}^{\alpha''\alpha'}\right).
\end{equation}
Rearranging the expression \eqref{24} we have
\begin{equation}\label{25}
   c_{1n}^\alpha \left( Z_{1n}^\alpha + Z_{2n}^\alpha \right) r_{n}^{\alpha\alpha'} + \sum_{m, \alpha''} f_{nm}^{\alpha\alpha''}  c_{1m}^{\alpha''} r_{m}^{\alpha''\alpha'}=
   c_{1n}^\alpha\delta_{n,0}\delta_{\alpha,\alpha'} \left( Z_{1n}^\alpha - Z_{2n}^\alpha \right) - \sum_{\alpha''} f_{n0}^{\alpha\alpha''}  c_{1}^{\alpha''} \delta_{\alpha''\alpha'}
\end{equation}
Expanding the expression \eqref{25} with the polarization indexes $\alpha$, $\alpha'$, and $\alpha''$, we obtain the infinite system of equations for the
reflected waves' amplitudes $r_{m}^{\alpha\alpha'}$ which can be presented as
\begin{equation}\label{27}
    \sum_m \begin{pmatrix} D_{nm}^{ss} && D_{nm}^{sp} \\D_{nm}^{ps} && D_{nm}^{pp} \end{pmatrix}
    \begin{pmatrix} r_{m}^{ss} & r_{m}^{sp} \\ r_{m}^{ps} & r_{m}^{pp} \end{pmatrix} =
    \begin{pmatrix} V_{n}^{ss} & V_{n}^{sp} \\ V_{n}^{ps} & V_{n}^{pp} \end{pmatrix},
\end{equation}
where
\begin{equation}\label{28}
    \begin{split}
        &D_{nm}^{ss} =  c_{1m}^{s} \left[ \delta_{nm}  \left( Z_{1n}^s + Z_{2n}^s \right) + f_{nm}^{ss} \right], \\
       & D_{nm}^{sp} = c_{1m}^p f_{nm}^{sp}, \\
        &D_{nm}^{ps} = c_{1m}^s f_{nm}^{ps}, \\
        &D_{nm}^{pp} = c_{1n}^{p} \left[ \delta_{nm}  \left( Z_{1n}^p + Z_{2n}^p \right)  + f_{nm}^{pp} \right].
    \end{split}
\end{equation}

with
\begin{equation}\label{29}
    \begin{split}
        &V_{n}^{ss} =  c_{1}^s \left[ \delta_{n,0}\left( Z_{1n}^s - Z_{2n}^s \right) - f_{n0}^{ss} \right], \\
        &V_{n}^{sp} = - D_{n0}^{sp}, \\
        &V_{n}^{ps} = - D_{n0}^{ps}, \\
        &V_{n}^{pp} = c_{1}^p \left[ \delta_{n,0} \left( Z_{1n}^p - Z_{2n}^p \right) - f_{n0}^{pp} \right],
    \end{split}
\end{equation}
or in a compact form
\begin{equation}\label{30}
   \sum_m \widehat{D}_{nm} \widehat{r}_{m} = \widehat{V}_n.
\end{equation}
Then, the matrix $\widehat{r}_{m} $ containing the reflected waves' amplitudes is
\begin{equation}\label{30a}
   \widehat{r}_{m} = \sum_n [\widehat{D}^{-1}]_{nm} \widehat{V}_n.
\end{equation}

\subsection{Diffraction efficiencies}
By definition, the diffraction efficiencies of reflected and transmitted waves,  $R_n$  and $T_n$, are as follows,
\begin{equation}\label{31}
   R_n = \frac{\text{Re}~q_{1n}}{q_1} \frac{|\mathbf{E}^r_n|^2}{|\mathbf{E}^i|^2}, \quad
   T_n = \frac{\text{Re}~q_{2n}}{q_1} \frac{|\mathbf{E}^t_n|^2}{|\mathbf{E}^i|^2},
\end{equation}
where
\begin{equation}\label{32}
\begin{split}
   &\mathbf{E}^i = E^{i,s} \mathbf{s}^i + E^{i,p} \mathbf{p}^i , \\
   &\mathbf{E}^r_n = \sum_{\alpha} \left\{ r_{n}^{s\alpha} E^{i, \alpha} \mathbf{s}^r_n +  r_{n}^{p\alpha} E^{i, \alpha} \mathbf{p}^r_n \right\}, \\
   &\mathbf{E}^t_n = \sum_{\alpha} \left\{ t_{n}^{s\alpha} E^{i, \alpha}  \mathbf{s}^t_n + t_{n}^{p\alpha} E^{i, \alpha}  \mathbf{p}^t_n \right\}.
\end{split}
\end{equation}
From \eqref{32} we have
\begin{equation}\label{33}
\begin{split}
   &|\mathbf{E}^i|^2 = |E^{i,s}|^2 +  |E^{i,p}|^2, \\
   &|\mathbf{E}^r_n|^2 = \sum_{\alpha = s,p} |r_{n}^{\alpha s} E^{i, s} + r_{n}^{\alpha p} E^{i, p}|^2, \\
   &|\mathbf{E}^t_n|^2 = \sum_{\alpha = s,p} |t_{n}^{\alpha s} E^{i, s} + t_{n}^{\alpha p} E^{i, p}|^2.
\end{split}
\end{equation}
Consequently, for the s-polarized wave the diffraction efficiencies are
\begin{equation}\label{34}
    R_n^s = \frac{\text{Re}~q_{1n}}{q_1} \left( |r_{n}^{ss}|^2  +  | r_{n}^{ps}|^2 \right), \quad
    T_n^s = \frac{\text{Re}~q_{2n}}{q_1}  \left( |t_{n}^{ss}|^2  +  | t_{n}^{ps}|^2 \right),
\end{equation}
and for the p-polarized wave
\begin{equation}\label{35}
    R_n^p = \frac{\text{Re}~q_{1n}}{q_1} \left( |r_{n}^{sp}|^2  +  | r_{n}^{pp}|^2 \right), \quad
    T_n^p = \frac{\text{Re}~q_{2n}}{q_1}  \left( |t_{n}^{sp}|^2  +  | t_{n}^{pp}|^2 \right).
\end{equation}

\section{Continuous graphene sheet}
In case of the continuous graphene sheet the problem becomes substantially simplified. Namely, from (\ref{21}) it follows
\begin{equation}\label{i1}
    \begin{split}
        &f^{ss} =  \alpha_1, \quad
        f^{sp} = - \alpha_2, \\
        &f^{ps} =  \alpha_2, \quad
        f^{pp} = \alpha_1.
    \end{split}
\end{equation}
where
\begin{equation}\label{i1a}
   \alpha_1 = \frac{4\pi}{c}\sigma_1, \quad \alpha_2 = \frac{4\pi}{c}\sigma_2,
\end{equation}
so that the system (\ref{30})
\begin{equation}\label{i2}
    \left( \begin{matrix} D^{ss} && D^{sp} \\  D^{ps} && D^{pp} \end{matrix} \right)
    \left( \begin{matrix} r^{ss} & r^{sp} \\ r^{ps} & r^{pp} \end{matrix}\right) =
    \left( \begin{matrix} V^{ss} & V^{sp} \\  V^{ps} & V^{pp} \end{matrix}\right),
\end{equation}
becomes
\begin{equation}\label{i3}
    \begin{split}
        &D^{ss} = c_{1}^{s}  ( Z_{1}^s + Z_{2}^s  + 2  \alpha_1 ), \\
        &D^{sp} = - c_{1}^p \alpha_2, \\
        &D^{ps} =  c_{1}^s \alpha_2, \\
        &D^{pp} = c_{1}^{p} (  Z_{1}^p + Z_{2}^p   + 2 \alpha_1 ).
    \end{split}
\end{equation}
\begin{equation}\label{i4}
    \begin{split}
        &V^{ss} =  c_{1}^s (  Z_{1}^s - Z_{2}^s  -2  \alpha_1 ), \\
        &V^{sp} =  c_{1}^p  \alpha_2, \\
        &V^{ps} = - c_{1}^s \alpha_2, \\
        &V^{pp} = c_{1}^p (  Z_{1}^p - Z_{2}^p - 2  \alpha_1 ).
    \end{split}
\end{equation}
Here $c_{2}^\alpha \equiv c_{20}^\alpha$, $Z_{2}^\alpha \equiv Z_{20}^\alpha$. Taking into account that $c_1^s =1$,
the determinant of the matrix $\widehat{D}$ in (\ref{i2}) is
\begin{equation}\label{i5}
    \Delta = c_{1}^p \left[ ( Z_{1}^s + Z_{2}^s  + \alpha_1) (  Z_{1}^p + Z_{2}^p   + \alpha_1 ) + \alpha_2^2 \right].
\end{equation}
The corresponding cofactors are
\begin{equation}\label{i6}
    \begin{split}
        &\Delta_{ss} = c_{1}^p \left[ ( Z_{1}^s - Z_{2}^s  -  \alpha_1) (  Z_{1}^p + Z_{2}^p   + \alpha_1 ) - \alpha_2^2 \right], \\
        &\Delta_{sp} = (c_{1}^p)^2 \alpha_2  Z_{1}^p, \\
        &\Delta_{ps} = - \alpha_2 Z_{1}^s , \\
        &\Delta_{pp} = c_{1}^p \left[ ( Z_{1}^s + Z_{2}^s +  \alpha_1) (  Z_{1}^p - Z_{2}^p   - \alpha_1 ) - \alpha_2^2 \right], \\
    \end{split}
\end{equation}
Then, the reflection coefficients are
\begin{equation}\label{i7}
    r^{\alpha\alpha'} = \frac{\Delta_{\alpha\alpha'}}{\Delta},
\end{equation}
or explicitly,
\begin{equation}\label{i8}
    \begin{split}
        &r^{ss} = \frac{( Z_{1}^s - Z_{2}^s  -  \alpha_1) (  Z_{1}^p + Z_{2}^p   + \alpha_1 ) - \alpha_2^2 }{( Z_{1}^s + Z_{2}^s  + \alpha_1) (  Z_{1}^p + Z_{2}^p   + \alpha_1 ) + \alpha_2^2}, \quad\quad
        r^{sp} = \frac{\alpha_2 c_{1}^p Z_{1}^p}{( Z_{1}^s + Z_{2}^s  + \alpha_1) (  Z_{1}^p + Z_{2}^p   + \alpha_1 ) + \alpha_2^2}, \\
        &r^{ps} = \frac{- \alpha_2 Z_{1}^s}{c_{1}^p \left[  ( Z_{1}^s + Z_{2}^s  + \alpha_1) (  Z_{1}^p + Z_{2}^p   + \alpha_1 ) + \alpha_2^2 \right]} , \quad
        r^{pp} = \frac{( Z_{1}^s + Z_{2}^s +  \alpha_1) (  Z_{1}^p - Z_{2}^p   - \alpha_1 ) - \alpha_2^2}{ ( Z_{1}^s + Z_{2}^s  + \alpha_1) (  Z_{1}^p + Z_{2}^p   + \alpha_1 ) + \alpha_2^2 }.
    \end{split}
\end{equation}
From the expression \eqref{23} it follows that
\begin{equation}\label{i9}
    t^{\alpha\alpha'} = \frac{c_{1}^\alpha}{c_{2}^\alpha}\left(\delta_{\alpha\alpha'} + r^{\alpha\alpha'}\right),
\end{equation}
or explicitly,
\begin{equation}\label{i10}
    \begin{split}
        &t^{ss} = \frac{2Z_1^s(  Z_{1}^p + Z_{2}^p   + \alpha_1 ) }{( Z_{1}^s + Z_{2}^s  + \alpha_1) (  Z_{1}^p + Z_{2}^p   + \alpha_1 ) + \alpha_2^2}, \quad \quad
        t^{sp} = \frac{2 \alpha_2 c_{1}^p Z_{1}^p}{( Z_{1}^s + Z_{2}^s  + \alpha_1) (  Z_{1}^p + Z_{2}^p   + \alpha_1 ) + \alpha_2^2}, \\
        &t^{ps} = \frac{- 2 \alpha_2 Z_{1}^s}{c_{2}^p \left[  ( Z_{1}^s + Z_{2}^s  + \alpha_1) (  Z_{1}^p + Z_{2}^p   + \alpha_1 ) + \alpha_2^2 \right]} , \quad
        t^{pp} = \frac{2 c_1^p Z_1^p ( Z_{1}^s + Z_{2}^s  + \alpha_1)}{ c_{2}^p \left[  ( Z_{1}^s + Z_{2}^s  + \alpha_1) (  Z_{1}^p + Z_{2}^p   + \alpha_1 ) + \alpha_2^2 \right] }. \\
    \end{split}
\end{equation}

\subsection{Free-standing graphene sheet}
For the free-standing graphene $\varepsilon_1 = \varepsilon_2 = 1$, and thus $q_1 = q_2 = q$, $c_1^p = c_2^p = c^p$,  $Z_1^\alpha = Z_2^\alpha = Z^\alpha$.
Then, the reflection coefficients are
\begin{equation}\label{i11}
    \begin{split}
        &r^{ss} = -\frac{ \alpha_1 (  Z^p +  \alpha_1 ) + \alpha_2^2 }{( Z^s  + \alpha_1) (  Z^p + \alpha_1 ) + \alpha_2^2}, \quad \quad
        r^{sp} = \frac{\alpha_2 c^p Z^p}{( Z^s + \alpha_1) (  Z^p  + \alpha_1 ) + \alpha_2^2}, \\
        &r^{ps} = \frac{- \alpha_2 Z^s}{c^p \left[  ( Z^s + \alpha_1) (  Z^p + \alpha_1 ) + \alpha_2^2 \right]} , \quad
        r^{pp} = - \frac{ \alpha_1 ( Z^s + \alpha_1)  + \alpha_2^2}{ ( Z^s + \alpha_1) (  Z^p + \alpha_1 ) + \alpha_2^2 }.
    \end{split}
\end{equation}
while the transmission coefficients have the expression
\begin{equation}\label{i12}
    \begin{split}
        &t^{ss} = \frac{Z^s(  Z^p + \alpha_1 ) }{( Z^s + \alpha_1) (  Z^p + \alpha_1 ) + \alpha_2^2},  \quad \quad \quad
        t^{sp} = \frac{ \alpha_2 c^p Z^p}{( Z^s  + \alpha_1) (  Z^p + \alpha_1 ) + \alpha_2^2}, \\
        &t^{ps} = \frac{- \alpha_2 Z^s}{c^p \left[  ( Z^s + \alpha_1) (  Z^p + \alpha_1 ) + \alpha_2^2 \right]} , \quad
        t^{pp} = \frac{Z^p ( Z^s  + \alpha_1)}{ \left[  ( Z^s + \alpha_1) (  Z^p + \alpha_1 ) + \alpha_2^2 \right] }.
    \end{split}
\end{equation}
\subsection{Normal incidence}
If the wave is normally incident on a free-standing continuous graphene sheet, $Z^s = Z^p =1$, $c^s = c^p = 1$ and thus:
\begin{equation}\label{i11}
    \begin{split}
        &r^{ss} = -\frac{ \alpha_1 (  1 +  \alpha_1) + \alpha_2^2 }{( 1  + \alpha_1)^2 + \alpha_2^2}, \quad \quad
        r^{sp} = \frac{\alpha_2}{( 1 + \alpha_1)^2 + \alpha_2^2}, \\
        &r^{ps} = - r^{sp}, \quad \quad \quad \quad \quad \quad \quad
        r^{pp} = r^{ss}.
    \end{split}
\end{equation}
\begin{equation}\label{ni12}
    \begin{split}
        &t^{ss} = \frac{1 + \alpha_1 }{( 1 + \alpha_1) ^2 + \alpha_2^2},  \quad \quad \quad
        t^{sp} = \frac{ \alpha_2 }{( 1  + \alpha_1)^2 + \alpha_2^2}, \\
        &t^{ps} = - t^{sp}, \quad \quad \quad \quad \quad \quad \quad
        t^{pp} = t^{ss}.
    \end{split}
\end{equation}
From \eqref{ni12} we can find the maximal value for cross-polarization coefficients, $|t^{sp}|^2$, $|t^{ps}|^2$, $|r^{sp}|^2$, $|r^{ps}|^2$. Let $a = 4 e^2 / (\hbar c) \equiv 2\alpha$,
where $\alpha\simeq 137^{-1}$ is a fine-structure constant. Then, substituting the explicit expressions for $\alpha_1$ and $\alpha_2$ into the expression for $t^{sp}$ we obtain
\begin{equation}\label{ni13}
     t^{sp} = \frac{a\beta}{\Omega^2-\beta^2 - (a+\gamma)^2 + 2i\Omega(a+\gamma)},
\end{equation}
and thus
\begin{equation}\label{ni14}
     |t^{sp}|^2 = \frac{a^2\beta^2}{(\Omega^2-\beta^2 - (a+\gamma)^2)^2 + 4\Omega^2 (a+\gamma)^2}.
\end{equation}
This expression reaches its maximum at
\begin{equation}\label{ni15}
     \Omega_0^2 = \beta^2 - (a+\gamma)^2,
\end{equation}
i.e. not strictly at the cyclotron resonance $\Omega = \beta$, but slightly shifted to lower frequencies.
For not very low cyclotron frequencies we can rewrite \eqref{ni15} as
\begin{equation}\label{ni16}
    \Omega_0 \simeq \beta  - \frac{(a+\gamma)^2}{2\beta},
\end{equation}
and in the first-order approximation $ \Omega_0 \simeq \beta$.
Substituting \eqref{ni15} back to the \eqref{ni14} we obtain
\begin{equation}\label{ni17}
     |t^{sp}|_{\text{max}}^2 = |t^{ps}|_{\text{max}}^2 =  |r^{sp}|_{\text{max}}^2 = |r^{ps}|_{\text{max}}^2 = \frac{a^2}{4(a+\gamma)^2}.
\end{equation}
If losses are absent, i.e. $\gamma \rightarrow 0$, we have
\begin{equation}\label{ni18}
     |t^{sp}|_{\text{max}}^2 = |t^{ps}|_{\text{max}}^2 =  |r^{sp}|_{\text{max}}^2 = |r^{ps}|_{\text{max}}^2 =  \frac{1}{4}.
\end{equation}

We recall that this condition has been found by an explicit calculation for a translationally-invariant graphene sheet. However, there is a more general proof for the statement that, in an infinitesimally thin film, the squared modulus of any cross-polarization zero-order scattering coefficient cannot exceed $1/4$.
We start by the following condition, which arises from current conservation:
\begin{equation}\label{ni19}
     \sum_{\alpha}|t^{\alpha\alpha'}|^2 + \sum_{\alpha}|r^{\alpha\alpha'}|^2 \leq 1.
\end{equation}
From Eq.~\eqref{i9}, for free-standing graphene in case of normal incidence we have
\begin{equation}\label{ni20}
    t^{\alpha\alpha'} = \delta_{\alpha\alpha'} + r^{\alpha\alpha'}.
\end{equation}
Substituting Eq.~\eqref{ni20} into Eq.~\eqref{ni19} we obtain
\begin{equation}\label{ni21}
    \sum_{\alpha}|\delta_{\alpha\alpha'} + r^{\alpha\alpha'}|^2 + \sum_{\alpha}|r^{\alpha\alpha'}|^2 \leq 1.
\end{equation}
If the incident wave is $p$-polarized ($\alpha' = p$) from Eq.~\eqref{ni21} we have
\begin{equation}\label{ni22}
   |1 + r^{pp}|^2 + |r^{sp}|^2 + |r^{sp}|^2 + |r^{pp}|^2 \leq  1,
\end{equation}
from where
\begin{equation}\label{ni23}
    2|r^{sp}|^2   \leq  1 - |1 + r^{pp}|^2 - |r^{pp}|^2.
\end{equation}
If we present $r^{pp}$ as a complex variable $z = x+iy$, we may introduce the function
\begin{equation}\label{ni24}
    f(z) = 1 - |1+z|^2 - |z|^2 = -2(x + x^2 + y^2).
\end{equation}
In the complex plane $(x,y)$ the function $f(z)$ has the maximum at the point $z_0 = -1/2$ where $f(z_0) = 1/2$.
Though $x$ and $y$ are not independent but $x = x(\omega)$, $y = y(\omega)$, and we cannot
guarantee that $f(x(\omega),y(\omega))$ can reach $1/2$, it still sets a restriction on $|r^{sp}|^2$: it cannot exceed $1/4$,
which is in accordance with the particular case stated in \eqref{ni18}.

\subsection{Metamaterial approach to the effective conductivity}
The relations \eqref{ni12} allow derivation of $\alpha_1$ and $\alpha_2$ from the any two independent field components $t^{s\alpha}$, $t^{p\alpha}$.
Let it be $t^{sp}$ and $t^{pp}$, then
 \begin{equation}\label{m1}
     (t^{sp})^2 + (t^{pp})^2 = \frac{1}{( 1  + \alpha_1)^2 + \alpha_2^2},
\end{equation}
and, therefore, from \eqref{ni12} it follows that
\begin{equation}\label{m2}
    \alpha_1 = \frac{t^{pp}}{(t^{sp})^2 + (t^{pp})^2} - 1, \quad \quad  \alpha_2 = \frac{t^{sp}}{(t^{sp})^2 + (t^{pp})^2}.
\end{equation}
Treating the ribbons array as a continuous layer with a certain effective conductivity, we may equate $t^{sp} = t_0^{sp}$ and $t^{pp} = t_0^{pp}$.
From the expressions \eqref{m2} we readily obtain $\widetilde\alpha_1$ and $\widetilde\alpha_2$:
\begin{equation}\label{m3}
    \widetilde\alpha_1 = \frac{t_0^{pp}}{(t_0^{sp})^2 + (t_0^{pp})^2} - 1, \quad \quad  \widetilde\alpha_2 = \frac{t_0^{sp}}{(t_0^{sp})^2 + (t_0^{pp})^2}.
\end{equation}
Finally, the two independent components of the effective conductivity are
\begin{equation}\label{m3}
    \widetilde\sigma_1 = \frac{c}{4\pi}\left[\frac{t_0^{pp}}{(t_0^{sp})^2 + (t_0^{pp})^2} - 1\right], \quad \quad
    \widetilde\sigma_2 =  \frac{c}{4\pi}\frac{t_0^{sp}}{(t_0^{sp})^2 + (t_0^{pp})^2}.
\end{equation}

\section{Polarization rotation}

\begin{figure}[h!]
    \includegraphics{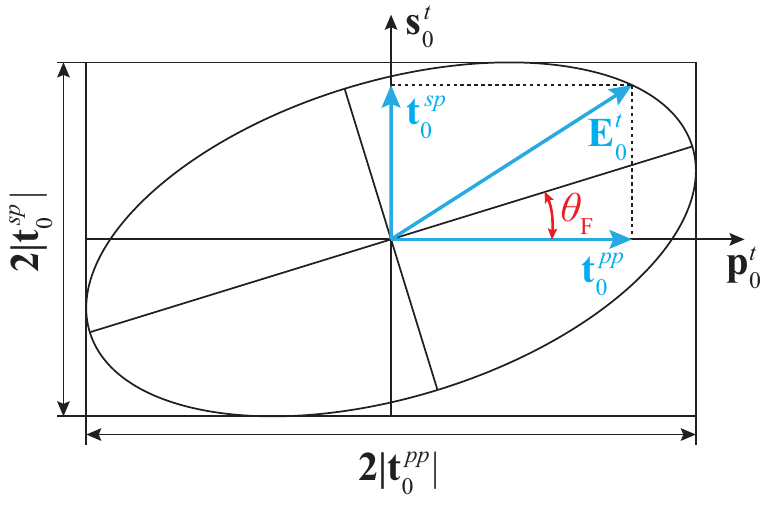}
    \caption{\label{supFig1} Decomposition of the zeroth-order transmitted wave into $s$ and $p$-polarized components.}
\end{figure}

In this section we will study the polarization rotation of the zeroth-order transmitted wave.
We are interested only in the case when the incident
wave is normally impinging the array of graphene ribbons. According to (\ref{32}), in case of the $p$-polarized incident wave with the unity amplitude ($ E^{i, p}=1$)
the amplitude of the zeroth-order transmitted wave is
\begin{equation}\label{p1}
   \mathbf{E}^t_0 = t_{0}^{pp} \mathbf{p}^t_0 + t_{0}^{sp} \mathbf{s}^t_0 ,
\end{equation}
which means that after passing through the ribbons array in a non-zero magnetic field, the zeroth-order transmitted wave
becomes elliptically polarized, see Fig.~\ref{supFig1}.

In terms of linearly polarized waves the polarization rotation angle $\theta_{\text{F}}$ is given by the expression:
\begin{equation}\label{p3}
   \theta_{\text{F}} = - \frac{1}{2} \arctan (e_1, e_2),
\end{equation}
with
\begin{equation}\label{p3_1}
        e_1 = 2 \Re (t_{0}^{pp} t_{0}^{sp, \ast}), \quad
        e_2 = |t_{0}^{pp}|^2-|t_{0}^{sp}|^2.
\end{equation}
Note that here $ \arctan(e_1, e_2)$ is a four quadrant arctangent function, $\arctan(e_1, e_2) = \arctan(e_2/e_1)$
which takes into account not the final sign of the quotient $e_2/e_1$, but the signs of $e_1$ and $e_2$ separately.

Let $A = 1+\widetilde\alpha_1$,  and $B = \widetilde\alpha_2$, then from \eqref{ni12} it follows that
 \begin{equation}\label{p4}
    t_0^{pp} = \frac{A }{A ^2 + B^2} = \frac{A(A^2+B^2)^\ast}{|A^2+B^2|^2},
\end{equation}
 \begin{equation}\label{p5}
    t_0^{sp,\ast} = \left( \frac{B }{A ^2 + B^2}\right)^\ast =  \frac{ B^\ast (A^2+B^2) }{|A^2 +B^2|^2}.
\end{equation}
Then, substituting \eqref{p4} and \eqref{p5} into \eqref{p3} we obtain
\begin{equation}\label{p6}
   \theta_{\text{F}} = - \frac{1}{2} \arctan \frac{2 \Re (AB^\ast)}{|A|^2-|B|^2} =
   -  \frac{1}{2} \arctan \frac{2 \Re (\widetilde\alpha_2^\ast + \widetilde\alpha_1\widetilde\alpha_2^\ast)}{| 1+\widetilde\alpha_1|^2-| \widetilde\alpha_2|^2}.
\end{equation}
In case of a low conductivity, i.e. when $|\alpha_1|, |\alpha_2| \ll 1$, we can keep only the linear in the conductivity terms and obtain the
widely adopted expression for the Faraday rotation angle
\begin{equation}\label{p7}
   \theta_{\text{F}} \simeq - \Re \widetilde\alpha_2 = - \frac{4\pi}{c} \Re \widetilde\sigma_2.
\end{equation}

\subsection{Maximal Faraday rotation in continuous graphene}

In terms of circularly polarized transmitted waves
\begin{equation}\label{p8}
    t_0^{+} = \frac{t_0^{pp} - i t_0^{sp}}{2}, \quad
    t_0^{-} = \frac{t_0^{pp} + i t_0^{sp}}{2},
\end{equation}
the Faraday angle is defined as
\begin{equation}\label{p9}
    \theta_{\text{F}} =  \frac{1}{2} \arg \frac{t_0^{+}}{t_0^{-}} = \frac{1}{2} \arg \frac{t_0^{pp} - i t_0^{sp}}{t_0^{pp} + i t_0^{sp}}.
\end{equation}
From \eqref{p9} we can find the maximal possible Faraday rotation angle for a continuous graphene sheet. Let
\begin{equation}\label{p10}
    Z = \frac{t^{pp} - i t^{sp}}{t^{pp} + i t^{sp}},
\end{equation}
then, substituting here the expressions \eqref{ni12} we obtain
\begin{equation}\label{p11}
    Z = \frac{1 + \alpha_1 - i\alpha_2}{1 + \alpha_1 + i\alpha_2}.
\end{equation}
Plugging in the explicit expressions for $\alpha_1$, $\alpha_2$ (see \eqref{1a} and \eqref{i1a}) and employing the notation $a = 4 e^2 / (\hbar c) \equiv 2\alpha$, we obtain
\begin{equation}\label{p12}
    Z = \frac{(\Omega - \beta + i\gamma)(\Omega + \beta + i(a+\gamma))}{(\Omega + \beta + i\gamma)(\Omega - \beta + i(a+\gamma))}.
\end{equation}
In the limit $\gamma \rightarrow 0$
\begin{equation}\label{p13}
    Z = \frac{(\Omega - \beta)(\Omega + \beta + ia)}{(\Omega + \beta)(\Omega - \beta + ia)} =
    \frac{(\Omega - \beta)(\Omega^2-\beta^2+a^2 - 2ia\beta)}{(\Omega + \beta)((\Omega - \beta)^2 + a^2)}.
\end{equation}
Analytical analysis of the function $\arg(Z)$ is rather time-consuming, so we present the result:
\begin{equation}\label{p15}
    \theta_{\text{F},\min}= -\frac{1}{2}\arctan \frac{\beta}{\alpha}   \quad \text{at} \quad \omega = \omega_{\text{c}}+0,
\end{equation}
\begin{equation}\label{p16}
    \theta_{\text{F},\max} = \frac{1}{2} \left[\pi-\arctan \frac{\beta}{\alpha} \right]   \quad \text{at} \quad \omega = \omega_{\text{c}}-0,
\end{equation}
where $\alpha$ is the fine-structure constant.
It is seen that with growth of the magnetic field, both values tend to $\pi/4$. Interestingly,
with magnetic field decrease $\theta_{\text{F},\min}\rightarrow0$ whereas $\theta_{\text{F},\min}\rightarrow\pi/2$.
It then follows that if we apply a very low magnetic field to a high-quality graphene sheet and then illuminate it at the frequency $\omega < \omega_{\text{c}}$,
the transmitted wave's polarization will be strongly rotated with almost no losses (of course, if the conductivity model works at such frequencies).
At zero magnetic field no polarization rotation can be obtained because we cannot go below $\omega_{\text{c}}$.
This result seems erroneous, however the first work by Crassee {\it et. al.} does indeed show
strong a asymmetry of Faraday rotation at low magnetic fields.


\begin{thebibliography}{20}

\bibitem{Shamir2006}{Shamir2006, J. Optical Systems and Processes; SPIE, 2006.}

\bibitem{Hopster2005}{Hopster, H.;  Oepen, H. P. Magnetic Microscopy of Nanostructures; Springer, 2005.}

\bibitem{Crassee2010}{Crassee, I.; Levallois, J.; Walter, A. L.; Ostler, M.; Bostwick, A.; Rotenberg, E.; Seyller, T.; van der Marel, D.; Kuzmenko, A. B.
Giant Faraday rotation in single- and multilayer graphene.
\emph{Nature Phys.} \textbf{2010}, 7, 48-51.}

\bibitem{Fallahi2012}{ Fallahi, A.; Perruisseau-Carrier, J.
Manipulation of giant Faraday rotation in graphene metasurfaces.
\emph{App. Phys. Lett.} \textbf{2012}, 101, 231605.}


%
%
%
%
%
%
%
%

\bibitem{Nikitin2012}{ Nikitin, A. Yu.; Guinea, F.; Garcia-Vidal, F. J.;  Mart\'{\i}n-Moreno, L.
Surface plasmon enhanced absorption and suppressed transmission in periodic arrays of graphene ribbons.
\emph{Phys. Rev. B} \textbf{2012}, 85, 081405(R) .}

\bibitem{Ju2011}{ Ju, L.; Geng, B.; Horng, J.; Girit, C.; Martin, M.; Hao, Z.; Bechtel, H. A.; Liang, X.; Zettl, A.; Shen, Y. R.; Wang, F.
Graphene plasmonics for tunable terahertz metamaterials.
\emph{Nature Nanotech.} \textbf{2011}, 6, 630-634.}

\bibitem{Vasic2013}{Vasi\'{c}, B.;  Isi\'{c}, G.; Gaji\'{c}, R.
Localized surface plasmon resonances in graphene ribbon arrays for sensing of dielectric environment at infrared frequencies.
\emph{J. Appl. Phys.} \textbf{2013}, 113, 013110.}

\bibitem{Nikitin2011}{Nikitin, A. Yu.; Guinea, F.; Garcia-Vidal, F. J.; Mart\'{\i}n-Moreno, L.
Edge and waveguide terahertz surface plasmon modes in graphene microribbons.
\emph{Phys. Rev. B} \textbf{2011},  84, 161407. }

\bibitem{Fang2012}{Fang, Z.; Wang, Y.; Liu, Z.; Schlather, A.; Ajayan, P. M.; Koppens, F. H. L.; Nordlander, P.;  Halas, N. J.
Plasmon-induced doping of graphene.
\emph{ACS nano} \textbf{2012}, 6, 10222-10228.}

\bibitem{Gao2012}{Gao, W.; Shu, J.; Qiu, C.; Xu, Q.
Excitation of plasmonic waves in graphene by guided-mode resonances.
\emph{ACS Nano} \textbf{2012}, 6, 7806-7813.}

\bibitem{Fang2013}{Fang, Z.; Thongrattanasiri, S.; Schlather, A.; Liu, Z.; Ma, L.; Wang, Y.; Ajayan, P. M.; Nordlander, P.; Halas, N. J.; Garc\'{\i}a de Abajo, F. J.
Gated Tunability and Hybridization of Localized Plasmons in Nanostructured Graphene.
\emph{ACS Nano} \textbf{2013}, 7, 2388-2395.}

\bibitem{Chiu1974}{ Chiu, K. W.; Quinn, J. J.
Plasma oscillations of a two-dimensional electron gas in a strong magnetic field.
\emph{Phys. Rev. B} \textbf{1974}, 9,  4724-4732.}

\bibitem{Kukushkin2006}{ Kukushkin, I. V.; Muravev, V. M.; Smet, J. H.; Hauser, M.; Dietsche, W.; von Klitzing, K.
Collective excitations in two-dimensional electron stripes: Transport and optical detection of resonant microwave absorption.
\emph{Phys. Rev. B} \textbf{2006}, 73, 113310.}

\bibitem{Bychkov2008}{ Bychkov, Y. A.; Martinez, G.
Magnetoplasmon excitations in graphene for filling factors $\text{v}\ll\text{6}$.
\emph{Phys. Rev. B} \textbf{2008}, 77, 125417.}

\bibitem{Berman2008}{ Berman, O. L.; Gumbs, G.; Lozovik, Yu. E.
Magnetoplasmons in layered graphene structures
\emph{Phys. Rev. B} \textbf{2008}, 78, 085401.}

\bibitem{Ferreira2012}{ Ferreira, A.; Peres, N. M. R.; Castro Neto, A. H.
Confined magneto-optical waves in graphene.
\emph{Phys. Rev. B} \textbf{2012}, 85, 205426.}

\bibitem{Sounas2011}{Sounas, D. L.; Caloz, C.
Edge surface modes in magnetically biased chemically doped graphene strips.
\emph{App. Phys. Lett.} \textbf{2011}, 99, 231902.}

\bibitem{Fischer2011}{ Fischer, A. M.; R\"{o}mer, R. A.; Dzyubenko, A. B.
Magnetoplasmons and SU(4) symmetry in graphene.
\emph{J. of Phys.: Conf. Ser.} \textbf{2011}, 286, 012054.}

\bibitem{Balev2011}{ Balev, O. G.; Vasilopoulos, P.; Frota, H. O.
Edge magnetoplasmons in wide armchair graphene ribbons.
\emph{Phys. Rev. B} \textbf{2011}, 84, 245406.}

\bibitem{GomezDiaz2012}{ G\'{o}mez-D\'{i}az, J. S.; Perruisseau-Carrier, J.
Propagation of hybrid transverse magnetic-transverse electric plasmons on magnetically biased graphene sheets.
\emph{J. App. Phys.} \textbf{2012}, 112, 124906.}

\bibitem{Giessen2013}{Chin, J. Y.; Steinle, T.; Wehlus, T.; Dregely ,D.; Weiss, T.; Belotelov, V. I.;
Stritzker, B.; Giessen, H. Nonreciprocal plasmonics enables giant enhancement of thin-film Faraday rotation,
\emph{Nat. Commun.} \textbf{2013},  4,  1599.}

\bibitem{BludovJAP012}{Bludov, Yu, V.; M. I. Vasilevskiy, M. I.; Peres, N. M. R.
Tunable graphene-based polarizer,
\emph{J. Appl. Phys.} \textbf{2012}, 112, 084320}

\bibitem{Ferreira2011}{Ferreira, A.; Viana-Gomes, J.; Bludov, Yu. V.; Pereira, V.; Peres, N. M. R.; Castro Neto, A. H.
Faraday effect in graphene enclosed in an optical cavity and the equation of motion method for the study of magneto-optical transport in solids.
\emph{Phys. Rev. B} \textbf{2011}, 84, 235410.}

\bibitem{Gusynin2009}{ Gusynin, V. P.; Sharapov, S. G.; Carbotte, J. P.
On the universal AC optical background in graphene.
\emph{New J. Phys.} \textbf{2009}, 11, 095013.}

\bibitem{Orlita2008}{ Orlita, M.; Faugeras, C.; Plochocka, P.; Neugebauer, P.; Martinez, G.; Maude, D. K.; Barra, A.-L.; Sprinkle, M.; Berger, C.; de Heer, W. A.; Potemski, M.
Approaching the Dirac point in high-mobility multilayer epitaxial graphene.
\emph{Phys. Rev. Lett.} \textbf{2008}, 101, 267601.}

\bibitem{Morozov2008}{ Morozov, S. V.; Novoselov, K. S.; Katsnelson, M. I.; Schedin, F.; Elias, D. C.; Jaszczak, J. A.; Geim, A. K.
Giant Intrinsic Carrier Mobilities in Graphene and Its Bilayer.
\emph{Phys. Rev. Lett.} \textbf{2008}, 100, 016602.}

\bibitem{Hwang2008}{ Hwang, E.; Das Sarma, S.
Acoustic phonon scattering limited carrier mobility in two-dimensional extrinsic graphene.
\emph{Phys. Rev. B} \textbf{2008}, 77, 115449.}

\bibitem{Chen2008}{ Chen, J.-H.; Jang, C.; Xiao, S.; Ishigami, M.; Fuhrer, M. S.
Intrinsic and extrinsic performance limits of graphene devices on $\text{SiO}_2$.
\emph{Nat. Nanotech.} \textbf{2008}, 3, 206-209.}

\bibitem{comsol}{We have used the implementation of the finite element method provided by the commercial software COMSOL
Multiphysics.}


\bibitem{Fialkovsky2009}{ Fialkovsky, I.V.; Vassilevich, D. V.
Parity-odd effects and polarization rotation in graphene.
\emph{J. of Phys. A} \textbf{2009}, 42, 442001.}

\bibitem{IBM012}{Yan,H.;  Low, T.;  Zhu, W.;  Wu, Y.;  Freitag, M.;  Li, X.;  Guinea, F.;  Avouris, P.;  Xia, F.; Damping pathways of mid-infrared plasmons in graphene nanostructures, \emph{Nature Photonics} \textbf{2013},  7, 394–399}






\bibitem{Alaee2012}{ Alaee, R.; Farhat, M.; Rockstuhl, C.; Lederer, F.
A perfect absorber made of a graphene micro-ribbon metamaterial.
\emph{Opt. Express} \textbf{2021}, 20, 28017.}


\bibitem{Eliasson1986}{ Eliasson, G.; Wu, J.-W.; Hawrylak, P.; Quinn, J. J.
Magnetoplasma modes of a spatially periodic two-dimensional electron gas
\emph{Solid State Commun.} \textbf{1986}, 60, 41-44.}

\bibitem{Mikhailov2005}{ Mikhailov, S.; Savostianova, M.
Microwave response of a two-dimensional electron stripe.
\emph{Phys. Rev. B} \textbf{2005}, 71, 035320.}






\bibitem{Demel1988}{ Demel, T.; Heitmann, D.; Grambow, P.; Ploog, K.
Far-infrared response of one-dimensional electronic systems in single- and two-layered quantum wires.
\emph{Phys. Rev. B} \textbf{1988}, 38, 12732-12735.}

\bibitem{Demel1991}{ Demel, T.; Heitmann, D.; Grambow, P.; Ploog, K.
One-dimensional plasmons in AlGaAs/GaAs quantum wires.
\emph{Phys. Rev. Lett.} \textbf{1991}, 66, 2657-2660.}

\bibitem{Zhao1994}{ Zhao, H. L.; Zhu, Y.; Wang, L.; Feng, S.
Magnetoplasmons in a quasi-one-dimensional quantum wire.
\emph{J. of Phys.} \textbf{1994}, 6, 1685.}


\bibitem{Kukushkin2005}{ Kukushkin, I. V.; Smet, J. H.; Kovalskii, V. A.; Gubarev, S. I.; von Klitzing, K.; Wegscheider, W.
Spectrum of one-dimensional plasmons in a single stripe of two-dimensional electrons.
\emph{Phys. Rev. B} \textbf{2005}, 72, 161317.}

\bibitem{Fedorych2009}{ Fedorych, O. M.; Studenikin, S. A.; Moreau, S.; Potemski, M.; Saku, T.; Hirayama, Y.
Microwave magnetoplasmon absorption by a 2DEG stripe.
\emph{Int. J. of Mod. Phys. B} \textbf{2009}, 23, 2698.}

\bibitem{Mikhailov2007}{ Mikhailov, S. A.; Savostianova, N. A.
Microwave Response of a Two-Dimensional Electron Stripe: Electrodynamics and The Influence of Contacts.
\emph{Int. J. of Mod. Phys. B} \textbf{2007}, 21, 1497-1501.}


\bibitem{Suko2012}{Thongrattanasiri, S.; Koppens, F. H. L.; Abajo, F. J. G.
Complete optical absorption in periodically patterned graphene.
\emph{Phys. Rev. Lett.} \textbf{2012}, 108, 047401.}










%
%
%







\end{thebibliography}
\end{document}